# The cellular ROS-scavenging function, a key factor determining the specific vulnerability of cancer cells to cold atmospheric plasma *in vitro*.

Dayun Yan[1*], Jonathan H. Sherman[2], Jerome Canady[3], Barry Trink[4], and Michael Keidar[1*]

[1]Department of Mechanical and Aerospace Engineering, The George Washington University, Science & Engineering Hall, 800 22nd Street, NW, Washington, DC 20052, USA

[2]Neurological Surgery, The George Washington University, Foggy Bottom South Pavilion, 22nd Street, NW, 7th Floor, Washington, DC 20037, USA

[3]Jerome Canady Research Institute for Advanced Biological and Technological Science, 6930 Carroll Avenue, 3rd floor Suite 300 Takoma Park, Maryland 20912, USA

[4]Department of Otolaryngology, Division of Head and Neck Cancer Research, School of Medicine, Johns Hopkins University, Baltimore, MD 21287, USA

*Corresponding authors: Dayun Yan ydy2012@gwmail.gwu.edu, Michael Keidar keidar@gwu.edu

**Abstract.** Cold atmospheric plasma (CAP) has shown its promising application in cancer treatment both *in vitro* and *in vivo*. However, the anti-cancer mechanism is still largely unknown. CAP may kill cancer cells via triggering the rise of intracellular ROS, DNA damage, mitochondrial damage, or cellular membrane damage. While, the specific vulnerability of cancer cells to CAP has been observed, the underlying mechanism of such cell-based specific vulnerability to CAP is completely unknown. Here, through the comparison of CAP treatment and $H_2O_2$ treatment on 10 different cancer cell lines *in vitro*, we observed that the $H_2O_2$ consumption speed by cancer cells was strongly correlated to the cytotoxicity of CAP treatment on cancer cells. Cancer cells that clear extracellular $H_2O_2$ more quickly are more resistant to the cytotoxicity of CAP treatment. This finding strongly indicates that the anti-oxidant system in cancer cells play a key role in the specific vulnerability of cancer cells to CAP treatment *in vitro*.

**Introduction.**

As a near-room temperature ionized gas, cold atmospheric plasma (CAP) has demonstrated its promising application in cancer treatment by causing the selective death of cancer cells *in vitro* [1–3]. The CAP treatment on several subcutaneous xenograft tumors and melanoma in mice has also demonstrated its potential clinical application [4–7]. The rise of intracellular ROS, DNA damage, mitochondrial damage, as well as apoptosis have been extensively observed in the CAP-treated cancer cell lines [8–10]. The increase of intracellular ROS may be due to the complicated intracellular pathways or the diffusion of extracellular ROS through the cellular membrane [11]. However, the exact underlying mechanism is still far from clear.

Cancer cells have shown specific vulnerabilities to CAP [12]. Understanding the vulnerability of cancer cells to CAP will provide key guidelines for its application in cancer treatment. Two general trends about the cancer cells' vulnerability to CAP treatment have been observed *in vitro*. First, the cancer cells expressing the p53 gene are more resistant to CAP treatment than p53 minus cancer cells [13]. p53, a key tumor suppressor gene, not only restricts abnormal cells via the induction of growth arrest or apoptosis, but also protects the genome from the oxidative damage of ROS such as $H_2O_2$ through regulating the intracellular redox state [14]. p53 is an upstream regulator of the expression of many anti-oxidant enzymes such as glutathione peroxidase (GPX), glutaredoxin 3 (Grx3), and manganese superoxide dismutase (MnSOD) [15]. In addition, cancer cells with a lower proliferation rate are more resistant to CAP than cancer cells with a higher proliferation rate [16]. This trend may be due to the general observation that the loss of p53 is a key step during tumorigenesis [17]. Tumors at a high tumorigenic stage are more likely to have lost p53 [17].

Despite the complicated interaction between CAP and cancer cells, the initial several hours after treatment has been found to be an important stage for the cytotoxicity of CAP. The anti-cancer ROS molecules in the extracellular medium are completely consumed by cells during this time period [12]. After the initial several hours, replacing the medium surrounding the cancer cells does not change the cytotoxicity of CAP [18]. Here, we first demonstrate that the $H_2O_2$ consumption speed of cancer cells after CAP treatment is a key factor determining the specific vulnerability of cancer cell lines to CAP. The higher $H_2O_2$ consumption speed of cancer cells during the initial 3 hours after CAP treatment, results in a less degree of cytotoxicity with CAP treatment. Cancer cells having the capacity to quickly clear the extracellular ROS are more likely to survive compared with other cells which consume the extracellular ROS more slowly.

**Methods and materials.**

**CAP device**. The CAP device used in this study is a typical CAP jet generator using helium as the carrying gas. The detailed description of this device has been illustrated in previous studies [12,19]. Briefly, a violet plasma jet is formed below the main discharge area between a central anode and an annular grounded cathode and flows out a quartz tube with a diameter of 4.5 mm. The discharge is driven by a 30 kHz alternating current voltage (3.02 kV). The flow rate of the carrying gas is about 4.7 L/min.

**Cell cultures**. Human pancreas ductal adenocarcinoma cell line (PANC-1) was purchased from American Type Culture Collection (ATCC). Other cell lines were donated by several labs at the George Washington University. These cells were all purchased from ATCC by the different labs. Human pancreatic adenocarcinoma cell line (PA-TU-8988T), human glioblastoma cell line (U87MG), as well as human lung carcinoma cell line (A549) were provided by Dr. Murad's lab. Human breast cancer cell lines (MDA-MB-231, MCF-7) were provided by Dr. Zhang's lab. Human ovarian carcinoma cell line (SK-OV-3), human ovarian carcinoma cell line (IGROV-1), human colorectal carcinoma cell line (HCT116), as well as human bone osteosarcoma cell line (U-2 OS) were provided by Dr. Zhu's lab. Murine melanoma cell line (B16F10) was provided by Dr. Sotomayor's lab. The medium used in the culture of B16F10 cells was composed of RPMI-1640 supplemented with 10% fetal bovine serum (Atlanta Biologicals, S11150) and 1% (v/v) penicillin and streptomycin solution (Life Technologies, 15140122). All other cells were cultured in DMEM supplemented with 10% (v/v) fetal bovine serum and 1% (v/v) penicillin and streptomycin solution. For each experiment, $3\times10^3$ cells were seeded per well on a 96-well plate (Falcon, 62406-081) and cultured 24 hours under standard culture conditions (a humidified, 37°C, 5% $CO_2$ environment) prior to CAP treatment.

**CAP treatment and $H_2O_2$ treatment on cancer cells.** Prior to CAP treatment, all entire medium used to culture cells overnight was removed. To perform the direct CAP treatment, the gap between the bottom of the 96-well plate and the CAP source was 3 cm. Subsequently, 100 μL of fresh DMEM or RPMI-1640 (only for B16F10 cells) was added to the cancer cells in the 96-well plate. The CAP jet was then used to vertically treat each well for 1 min, 2 min, or 3 min. $H_2O_2$-containing medium was made by adding 9.8 M $H_2O_2$ standard solution (216763, Sigma-Aldrich) in DMEM or RPMI-1640 (only for B16F10 cells). 100 μL of $H_2O_2$-containing medium was then added to the cancer cells. After direct CAP and $H_2O_2$ treatment, the cancer cells were cultured under standard conditions for 3 days prior to performing the cell viability assay. In all cases, the control group consisted of cancer cells grown in fresh DMEM without CAP or $H_2O_2$ treatment.

**Cell viability assay.** MTT (3-(4,5-Dimethyl-2-thiazol)-2,5-Diphenyl-2H-tetrazolium Bromide) assay was performed following the standard protocols provided by Sigma-Aldrich. The absorbance at 570 nm was measured by a H1 microplate reader (Hybrid Technology). The measured absorbance was processed to be a relative cell viability by the division between the data of the experimental group and the control group.

**Extracellular $H_2O_2$ assay**. The $H_2O_2$ concentration was measured using the Fluorimetric Hydrogen Peroxide Assay Kit (Sigma-Aldrich, MAK165-1KT) using standard protocols provided by Sigma-Aldrich. The fluorescence was measured by a H1 microplate reader (Hybrid Technology) at 540/590 nm. The final fluorescence was obtained by deducting the fluorescence of control group from the fluorescence of experimental group. The $H_2O_2$ concentration was obtained based on the standard curve.

**Measuring the $H_2O_2$ consumption speed by cancer cells.** The CAP-stimulated DMEM (PSM) was made by treating 8 mL DMEM in the well on a 6-well plate for 8 min. $H_2O_2$-containing DMEM was made as above. The same protocol was used on all cell lines. First, 100 μL of cells at a concentration of $6 \times 10^4$ cells/mL was seeded in each well. 3 wells were used for each test. Cells were then cultured for 24 hours under standard conditions. 100 μL of sample solution was added to the wells. After 3 hours, 50 μL of medium was collected and immediately transferred to a well on a black clear bottom 96-well plate (Falcon) and the $H_2O_2$ assay was then performed every hour in triplicate.

**Results and discussion.**

The initial several hours are the most important stage for determining the cytotoxicity of CAP on cancer cells [10,12,18]. Our previous studies have demonstrated that key reactive species such as $H_2O_2$ in the medium is completely consumed by glioblastoma cells in just 3 hours [10,12,18]. Here, we comprehensively compared the $H_2O_2$ consumption speeds of 10 cancer cell lines during their initial 3 hours cultured in the CAP-stimulated medium, which was used to quantify the ROS-scavenging ability of cancer cells. We measured the residual $H_2O_2$ in the medium surrounding the cells every hour after treatment for 3 hours. The relative residual $H_2O_2$ concentration was obtained by the division between the residual $H_2O_2$ concentration and the initial $H_2O_2$ generation in DMEM.

We found that the $H_2O_2$ consumption speed is cell specific. In the CAP-treated DMEM, B16F10 cells and SK-OV-3 cells consume $H_2O_2$ fastest among all 10 cell lines (Fig. 1a). B16F10 cells and SK-OV-3 cells clear all extracellular $H_2O_2$ in only 2 hours. U2 OS cells consume extracellular $H_2O_2$ the slowest (Fig. 1a). U2 OS cells only consume about 70% of extracellular $H_2O_2$ in three hours after CAP treatment. In addition, PA-TU-8998T, MCF-7 HCT116, and IGROV-1 cells have a similar but higher $H_2O_2$ consumption rates compared to U2 OS cells (Fig. 1a).

In addition, the 10 cancer cell lines showed specific vulnerability to direct CAP treatment. Due to the potential cell-based $H_2O_2$ generation during direct CAP treatment, the CAP device was used at a relatively low discharge voltage (3.02 kV). At such a low voltage, the cell-based $H_2O_2$ generation can be inhibited [20]. Thus, the initial reactive species input from CAP is the same among all cell lines. Among these cell lines, B16F10 and SK-OV-3 cells are most resistant to CAP treatment (Fig. 1b). A 3 min of CAP treatment led to only a 50% inhibition of cell viability in B16F10 cells compared with control. In contrast, U2OS, PA-TU-8998T, MCF-7, and HCT116 cells are the most vulnerable to direct CAP treatment (Fig. 1b). Clearly, these cancer cell lines have the least extracellular $H_2O_2$ consumption rates (Fig. 1a). The remaining cell lines generally follow this trend, in that the extracellular ROS scavenging ability of cancer cells is inversely proportional to their vulnerabilities to CAP treatment.

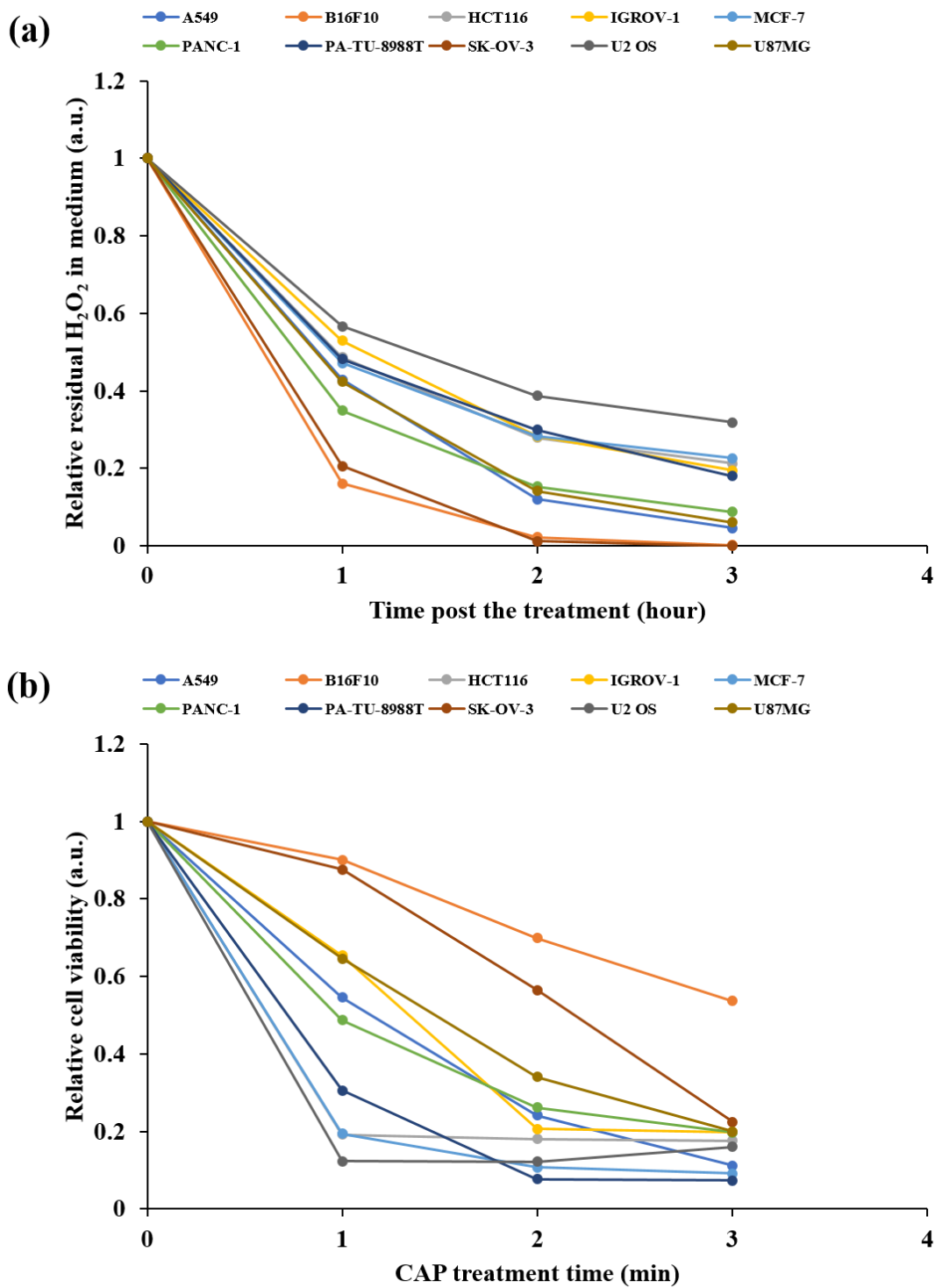

**Fig. 1. The effect of CAP treatment on 10 cancer cell lines.** (a) The evolution of $H_2O_2$ in the extracellular environment due to the consumption of cancer cells. $H_2O_2$ was generated by the CAP

treatment on the medium. (b) The cytotoxicity of direct CAP treatment on cancer cells grown on 96-well plates. The results are presented as the mean of three independent experiments performed in triplicate. The measured $H_2O_2$ concentration in the CAP-treated medium was $48.8 \pm 6.5$ μM. The original data with standard deviation are shown in supporting materials (Fig. S1 and S2).

This trend was preserved when cancer cell lines were grown in $H_2O_2$-containing medium. All 10 cancer cells showed nearly the same specific $H_2O_2$ consumption rates in the $H_2O_2$-containing medium as that observed in the CAP-stimulated medium (Fig. 2a). Similar, the vulnerability of cancer cells to $H_2O_2$ treatment is also inversely proportional to the $H_2O_2$ consumption rate of cancer cells (Fig. 2b). For example, B16F10 cells, simultaneously has the strongest $H_2O_2$-scavenging capacity and the strongest resistance to $H_2O_2$ treatment. As we observed in previous studies, however, CAP treatment cannot be regarded as a simple $H_2O_2$ treatment [12]. B16F10 cells are much more resistant to $H_2O_2$ treatment than all other cell lines include-ing SK-OV-3 cells (Fig. 2b). Instead, the vulnerability difference between B16F10 cells and SK-OV-3 cells during $H_2O_2$ treatment are much larger than that observed during CAP treatment. This difference may be due to the complicated ROS and RNS components generated in CAP treatment, which will never be generated by just a $H_2O_2$ treatment [21,22]. Nonetheless, the same trends observed in Fig. 1 and Fig. 2 clearly demonstrate that the ROS-scavenging potential of cancer cells may play key a role in the specific cytotoxicity of the extracellular ROS treatment of both CAP and $H_2O_2$ treatment.

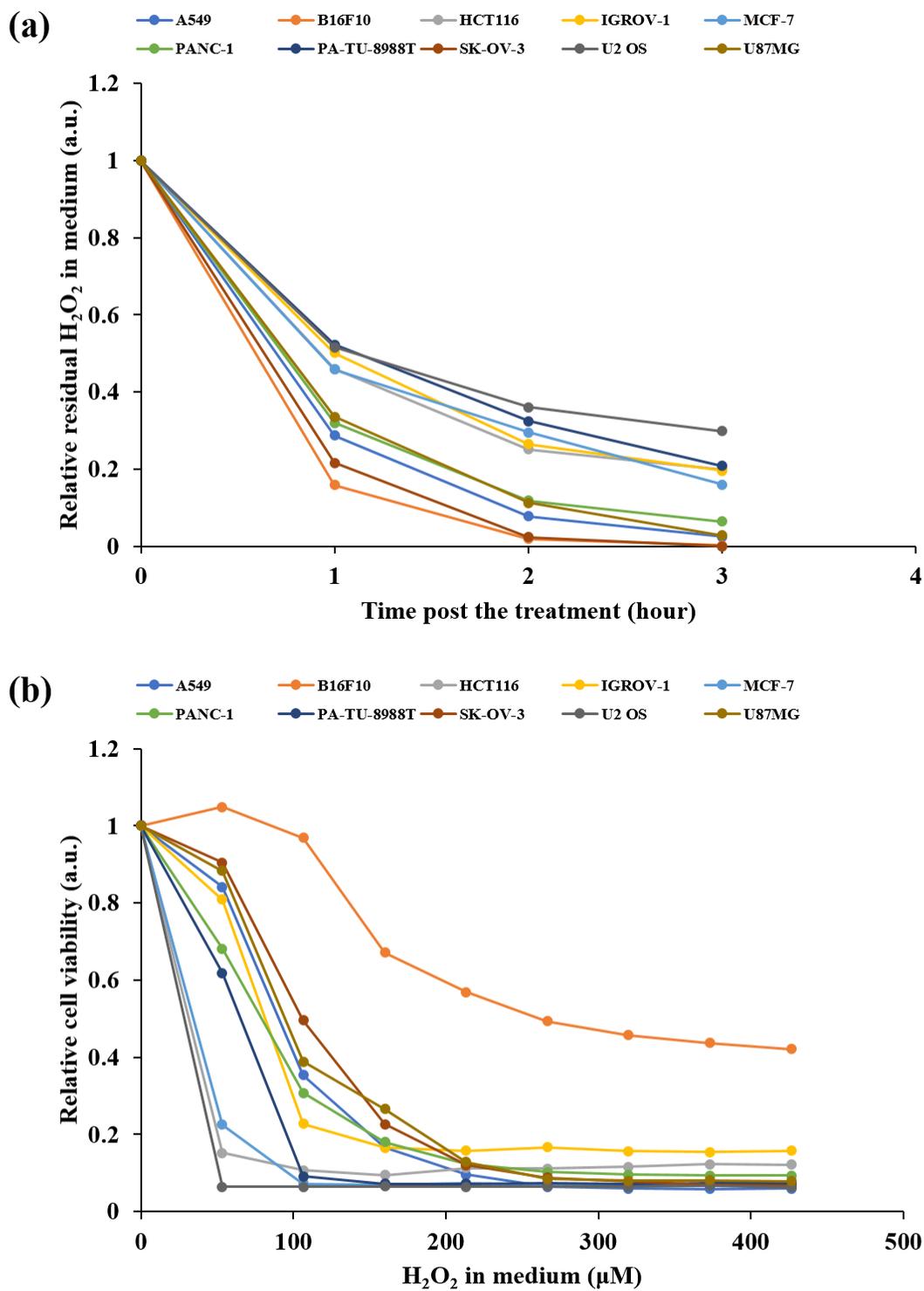

Fig. 2. The effect of H$_2$O$_2$ treatment on 10 cancer cell lines. (a) The evolution of H$_2$O$_2$ in the extracellular environment due to its consumption by cancer cells. The initial H$_2$O$_2$ concentration

was 53.3 ± 9.1 µM, which was the $H_2O_2$ concentration generated in the 8 mL medium after 8 min of CAP treatment. (b) The cytotoxicity of $H_2O_2$ treatment on cancer cells. The $H_2O_2$ concentrations shown here are the integral multiples of 53.3 µM. The results are presented as the mean of three independent experiments performed in triplicate. The original data with standard deviation are shown in supporting materials (Fig. S3 and S4).

The correlation between the $H_2O_2$ consumption potential of cancer cells and the cytotoxicity of CAP treatment or $H_2O_2$ treatment on cancer cells is summarized and shown in Fig. 3. The inversely proportional correlation between the $H_2O_2$ consumption rate and the cytotoxicity of CAP is more pronounced in the case of the CAP treatment (Fig. 3a) than that of the $H_2O_2$ treatment (Fig. 3b). Our finding provides a simple cellular marker to predict the cytotoxicity of CAP treatment on different cancer cells. To date, investigating the expression of p53 gene in cancer cells is the only general method to predict the specific cytotoxicity of CAP treatment. The cancer cells expressing p53 gene tend to be more resistant to CAP treatment compared to the cells without the p53 gene [13]. However, this strategy needs complicated biochemical analysis such as western blot and PCR. In contrast, our strategy needs only a measure of the extracellular $H_2O_2$ generation during the initial several hours including the first hour post CAP treatment. This is the first attempt to connect the previous observed complicated vulnerabilities of different cancer cells to CAP treatment with a clear but easily measurable cellular feature.

The $H_2O_2$ consumption speed of cancer cells may be the explanation at the cellular level for the correlation between the expression of p53 gene and the specific cytotoxicity of CAP treatment. p53 regulates the expression of the anti-oxidant system [15]. Thus, the vulnerability of cancer cells to CAP treatment may be significantly affected by the intracellular anti-oxidant system. For example, A549 and U87MG cells are known as peroxide-resistant cell lines [23]. The overexpression of the bcl-2 and the related bcl-xL protooncogene proteins and catalase may contribute to their $H_2O_2$-resistant feature through inhibiting apoptosis induced by oxidants and the scavenging intracellular $H_2O_2$, respectively [23]. The catalase activity is a major determinant of the cellular resistance to $H_2O_2$ toxicity [24]. The specific catalase expression levels in cancer cells may explain the correlation between the specific $H_2O_2$ consumption speed of cancer cells and the specific vulnerability of cancer cells to CAP treatment or $H_2O_2$ treatment. This explanation is consistent with our previous model that catalase may play an important role in the selective anti-cancer capacity of CAP, since cancer cells tend to express less catalase compared with their corresponding homologous normal cell lines in many cases [1].

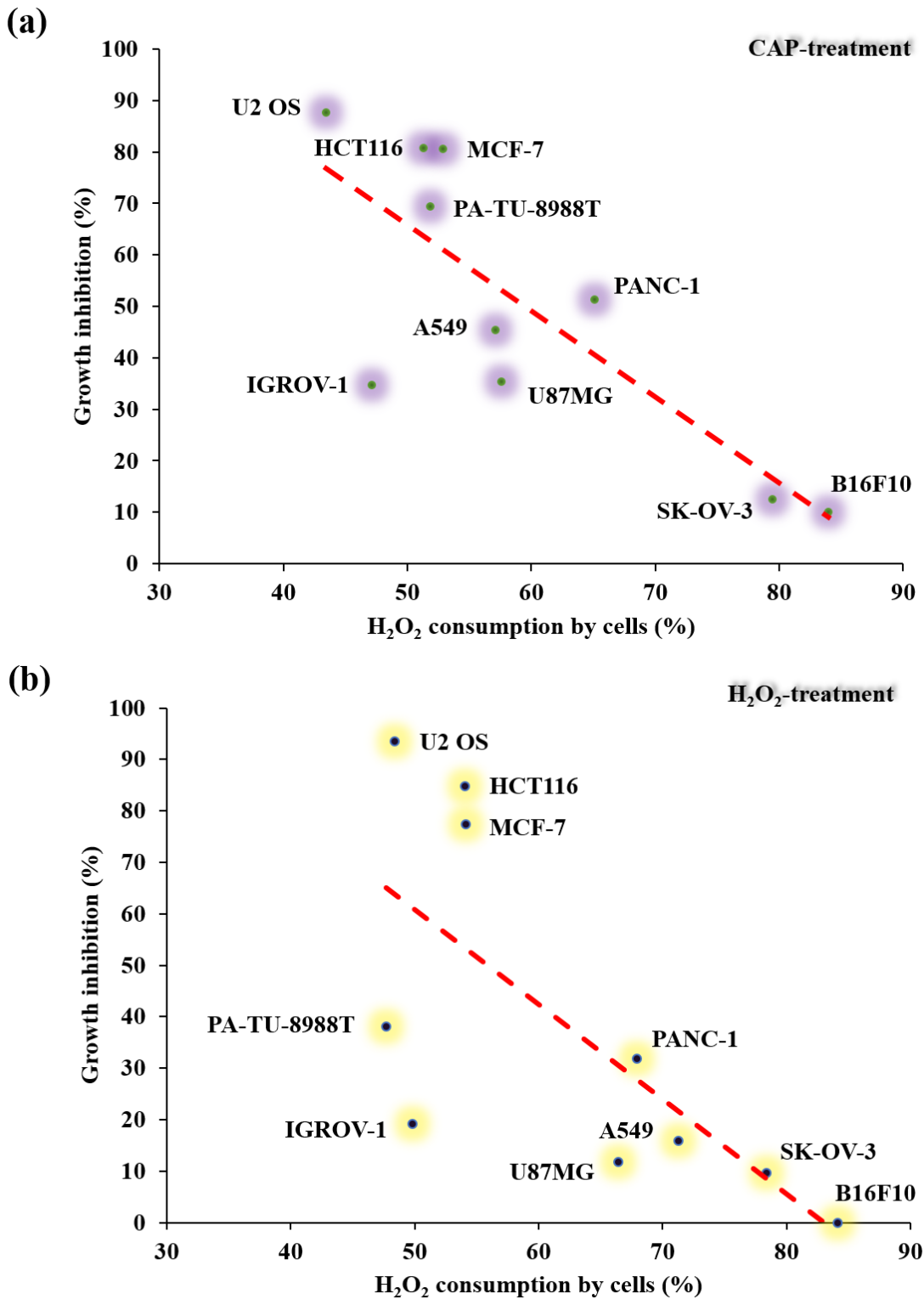

Fig. 3. The inversely proportional correlation between the $H_2O_2$ consumption rates of cancer cells and the growth inhibition effect of CAP treatment or $H_2O_2$ treatment on cancer cells.

(a) CAP treatment. (b) $H_2O_2$ treatment. The red lines are the trend lines drawn by Excel 2016. The $H_2O_2$ consumption rate and the growth inhibition rate were calculated based on the following formulas. The $H_2O_2$ consumption rate (%) = 100 x (1-the residual $H_2O_2$ concentration measured at the $1^{st}$ hour post the treatment). The residual $H_2O_2$ concentration was shown in Fig. 1a, and Fig. 2a. For CAP treatment, the growth inhibition rate (%) = 100 x (1 - relative cell viability [1 min of CAP treatment, Fig. 1b]). For $H_2O_2$ treatment, the growth inhibition rate (%) = 100 x (1 - relative cell viability [53.3 µM of $H_2O_2$ treatment, Fig. 2b]).

**Conclusions.**

The $H_2O_2$ consumption rate of cancer cells is an important cellular physiological marker to predict the cytotoxicity of CAP treatment or $H_2O_2$ treatment on cancer cell lines *in vitro*. The cancer cells which can clear the extracellular $H_2O_2$ at a faster rate tend to show stronger resistance to CAP treatment or $H_2O_2$ treatment. This trend firstly provides a simple cellular marker to correlate the vulnerability of cancer cells to CAP treatment. The measurement on the evolution of $H_2O_2$ during the initial several hours post CAP treatment can predict the final cytotoxicity of CAP treatment on cancer cells *in vitro*. The specific expression of intracellular anti-oxidant systems may contribute to such a unique correlation in cancer cells.

**Acknowledgements.** This work was supported by National Science Foundation, grant 1465061. This was is also supported in part by USPI Inc. Sincerely thanks for the cells donation from Prof. Murad, Prof. Zhu, Prof. Zhang, and Prof. Sotomayor at the George Washington University. Sincerely thanks for the help from Dr. Fengdong Cheng and Dr. Chi-Wei Chen on some cells' initial culture and seeding.